# Numerical Analysis of Flow Characteristics of An Atmospheric Plasma Torch

You-Jae Kim, J.-G. Han and Youn J. Kim[a]

[a]*Center for Advanced Plasma Surface Technology*
*Sungkyunkwan University*
*300 CheonCheon-dong, Suwon 440-746, KOREA*

**Abstract**

The atmospheric plasma is regarded as an effective method for surface treatments because it can reduce the period of process and does not need expensive vacuum apparatus. The performance of non-transferred plasma torches is significantly depended on jet flow characteristics out of the nozzle. In order to produce the high performance of a torch, the maximum discharge velocity near an annular gap in the torch should be maintained. Also, the compulsory swirl is being produced to gain the shape that can concentrate the plasma at the center of gas flow. Numerical analysis of two different mathematical models used for simulating plasma characteristics inside an atmospheric plasma torch is carried out. A qualitative comparison is made in this study to test the accuracy of these two different model predictions of an atmospheric plasma torch. Numerical investigations are carried out to examine the influence of different model assumptions on the resulting plasma characteristics. Significant variations in the results in terms of the plasma velocity and temperature are observed. These variations will influence the subsequent particle dynamics in the thermal spraying process. The uniformity of plasma distribution is investigated. For analyzing the swirl effects in the plenum chamber and the flow distribution, FVM (finite volume method) and a SIMPLE algorithm are used for solving the governing equations.

**Key Words:** plasma torch, swirl effect, spraying, plasma distribution

## 1. Introduction

In order to provide high quality for specialized applications and to aid in the development of the technology, many studies of electric arcs have been made. Non-transferred arc plasma torches have been used for producing thermal plasma jet in various industrial applications, such as atmospheric or low-pressure plasma spraying, thermal plasma waste treatment, plasma-assisted chemical vapor deposition, plasma preparation of ultra-fine powders, etc. Material processes by non-transferred torches are usually executed at the outside of the torch where the thermal plasma is exposed to an ambient gas environment

[1].

The majority of existing methods of using plasma torch generates the plasma only in a vacuum state. And due to the higher expense of vacuum apparatus and generator, the existing method is lack of economical efficiency and productivity. For resolving these problems, the velocity distribution is numerically analyzed and the size of flame is also experimentally photographed.

For stabilizing the plasma that is generated in using atmospheric plasma torch, flow behavior is the most important factor [2]. Also by concentrating the generated plasma in the center of jet flow, the uniformity for the surface treatment may be obtained. There are two kinds of methods to get these effects. Firstly, there is the wall-stabilization in which the plasma is concentrated to the center by decreasing the electrical conductivity through the cooling of the outer wall while surrounding the arc occurred space with cold metal wall. This is resulted in the stabilization of a flame and decrease the heat loss. Secondly, there is the vortex-stabilization in which the plasma is surrounded by swirl that is occurred by injecting the plasma gas. For the replacement of large-sized plasma torches, Kuo [3] studied in miniaturizing the existing plasma torches as shown in Fig. 1.

There are many researches to study the flow characteristics of plasma torch using some numerical models. In the present work, numerical analysis of two different mathematical models, constant heat source model (CHS) [4-6] and magneto-hydrodynamics model (MHD) [7], used for simulating the plasma characteristics of an atmospheric plasma torch is carried out to observe the discharge phenomenon by gas kinds that can influence the surface treatment of materials.

## 2. Numerical Analysis
### 2.1 Governing equations

In order to analyze the flow characteristics of a plasma torch, the governing equations for steady-state turbulent flows can be written as:

$$\frac{\partial}{\partial x}(\rho u \phi) + \frac{1}{r}\frac{\partial}{\partial r}(r \rho \upsilon \phi) = \frac{\partial}{\partial x}\left(\Gamma_\phi \frac{\partial \phi}{\partial x}\right) + \frac{1}{r}\frac{\partial}{\partial r}\left(r \Gamma_\phi \frac{\partial \phi}{\partial r}\right) + S_\phi \qquad (1)$$

where $\phi$ is the general variable, $\Gamma_\phi$ the corresponding diffusion coefficient, and $S_\phi$ the source term; $u$ and $\upsilon$ are the axial and radial velocity components, $\rho$ is the mass density, and $x$ and $r$ are the distances in the axial and radial directions, respectively. The terms of the governing equations are listed in Table 1. It is to be noted that in the calculation of the total energy deposited within the torch volume in the CHS model, we take into account the full nozzle volume due to the fact that the arc length for the given operating conditions is almost the same as the length of the torch nozzle. A few studies employed constant heat source (CHS) model by assuming a volume averaged source term instead of the Joule heating model in the energy equation. The major difference between the CHS and MHD models is reflected in their source terms of the momentum and energy equations, respectively. It is also noted that in the case of the MHD model we have additional force terms, due to the Lorentz force in the momentum equations, which are absent in the CHS model.

In view of the computational problems a few authors dealt the plasma generation inside the torch by a simplified approach which is known as the CHS model. In the CHS model, the local arc phenomenon inside the spray torch is replaced by a volume averaged source term in the energy equation to represent for the heating of primary gas by the electric arc. The energy source term is given by $S_E = (P_{out}/Torch\ Volume)$, where $P_{out} = \lambda V I$, denotes the torch power output ($kW$), $V$ is the arc voltage ($V$), $I$ is the arc current ($A$), and $\lambda$ is the torch efficiency factors respectively. However, the CHS model does not consider the influence of the Lorentz force that drives the gas inside the torch, which is reported to be significant especially in high intensity arcs.

In Table 1, variables $j_x$, $j_r$ and $B_\theta$ denote the radial, axial components of current density and induced magnetic field respectively and are computed from the electric potential $\zeta$ as:

$$j_x = -\sigma_\varepsilon \sqrt{\lambda} \frac{\partial \zeta}{\partial x}, \quad j_y = -\sigma_\varepsilon \sqrt{\lambda} \frac{\partial \zeta}{\partial y} \quad \text{and} \quad B_\theta = -\frac{\mu_0}{\sigma_\varepsilon} \int_0^r j_x \eta d\eta. \tag{2}$$

The effects of turbulence on the energy and momentum transports are estimated using the standard $k-\varepsilon$ turbulence model, which is suitable for turbulent kinetic energy $k$ and its dissipation rate $\varepsilon$. The constants appearing in the turbulent model are $C_{1\varepsilon} = 1.44$ and $C_{2\varepsilon} = 1.92$ [8].

The viscosity is assumed to apply the following Sutherland-law [9]:

$$\frac{\mu}{\mu_0} = \left(\frac{T}{T_0}\right)^{3/2} \frac{T_0 + S}{T + S} \tag{3}$$

where $\mu_0$ and $T_0$ are referenced values, and $S$ is an effective temperature, called Sutherland constant, which is a characteristic of the gas. The viscosity parameters of the Sutherland-law for gases are listed in Table 2. In the present study, incompressible gases with temperature dependent thermodynamic and transport properties are assumed to be in locally equilibrium. Different effects, namely ionization, radiation, electron condensation, etc on the plasma characteristics are not considered here.

**2.2 Model and numerical methods**

Calculation model is used in size of practical torch as shown in Fig. 1. The calculation domain is composed to 60mm×80mm. In order to elucidate the flow characteristics of plasma torch, two-dimensional grid system was made. The SIMPLE (Semi-Implicit Method for Pressure-Linked Equations) algorithm and hybrid scheme are also employed. In the present study, a solution is deemed converged when the mass imbalance in the continuity equation is less than $10^{-5}$. In addition, no-slip condition is used on the wall boundary. This is presumed that there is no mass flux on walls. In order to reduce a number of cells, the wall function was used. Also, Neumann condition that has zero gradient for all flow variables along streamline was applied on the outlet boundary, since the flow variables were difficult to know on the outlet boundary. For the numerical calculation, we used air, argon, nitrogen and helium gases as working fluids and 2 m/s as inlet velocity. Then the current and the power output of the torch are considered as 250A and 5kW, respectively.

## 3. Results and Discussion

The aim of the present study is to compare the numerical results of the discharge characteristics of the atmospheric plasma torch gained from MHD model with those of the CHS model.

Figure 2 shows the contours of axial velocity of the torch with swirl and without swirl effects. It is seen that the maximum velocity with swirl showed about 2 times that with the case of no-swirl.

Then the profiles of the centerline axial velocity in each gas species are shown in Fig. 3. Here x-axis represents the centerline of the torch and the origin means the end of the torch and the start of the discharge. The profiles of the axial velocity computed using the CHS and MHD models show large differences in their distribution. It is noted that the velocity distribution with the CHS model is aligned more uniformly than that of the MHD model. Then the CHS model shows lower velocity distribution than that of the MHD one, due to the absence of the Lorentz force assumptions in momentum equations. It is also seen that the maximum axial velocity of the MHD models is larger than that of the CHS model. This is due to the existence of high electrical field strength near the cathode in the MHD model, which is not taken into account in the CHS model.

As shown in Fig. 3, the velocity profile of nitrogen in the MHD model is similar to that of air along the centerline. And the velocity profiles of nitrogen and air in the MHD model from the origin to 0.04 m along the x-axis are higher than those of the other gas species. But the velocity profile of argon from 0.04 m to 0.08 m is higher than that of the other gas species. Results show that the velocity magnitude of helium is the smallest of them. These results represent the effect of the different viscosity parameters of the Sutherland-law in each gas species.

Figure 4 shows the influence of both the CHS and MHD models in each gas species on the distribution of the centerline axial discharge temperature from the exit of the torch. It is seen that the profiles of the discharge temperature computed using the CHS and MHD models show large differences in their distribution. The variations of temperature in the CHS model are smaller and more stable than those of the MHD model. And their respective temperature distributions in the MHD model are found significantly different, especially with steeper variations than those of the CHS model near the exit of the torch. This is due to the existence of high electric field strength inside the torch before discharge in the MHD model, which cannot be captured by the CHS model.

It is noted that the CHS model could be used to predict the plasma dynamics in each gas species and at high intensity arcs. However, it is suggested that the CHS model may be used only when the Lorentz force is negligible, which is true in the case of low intensity arcs ($I < 50\ A$) [10]. As we observe significant variations both in the discharge velocity and temperature computed through different models, between the CHS and MHD models, the resulting particle dynamics is expected to affect significantly.

## 4. Conclusions

The effect of the swirl flow is studied numerically in the atmospheric plasma torch. The comparison results between the MHD and CHS models are depicted. Due to lack of the experimental data inside the plasma torch, no direct comparisons have been made. But the discharge behaviors by gas species are studied experimentally to operate the highest plasma flame. The major conclusions are as follows:

1) The method of vortex-stabilization is acquired by the occurred swirl flow.
2) The MHD model is more realistic in high intensity arcs than the CHS model.

## Acknowledgement

The authors are grateful for the financial support provided by the Korea Science and Engineering Foundation through CAPST (Center for Advanced Plasma Surface Technology) at the Sungkyunkwan University.

**Table 1 Transport variables, coefficients and source terms**

| Equation | Process variables $\phi$ | Coefficient of diffusive term $\Gamma_\phi$ | Source terms $S_\phi$ | |
|---|---|---|---|---|
| | | | CHS model | MHD model |
| Mass | 1 | 0 | 0 | 0 |
| Momentum | $u$ | $\mu$ | $-\dfrac{\partial p}{\partial x}$ | $-\dfrac{\partial p}{\partial x} - j_r B_\theta \lambda$ |
| | $\upsilon$ | | $-\dfrac{\partial p}{\partial r}$ | $-\dfrac{\partial p}{\partial r} - j_x B_\theta \lambda$ |
| Energy | $T, \rho = \rho c_p$ | $k$ | $\dfrac{P_{out}}{Nozzle\ volume}$ | $\lambda \dfrac{j_x^2 + j_r^2}{\sigma_\varepsilon}$ |
| Electric potential | $\zeta, \rho = 0$ | $\sigma_\varepsilon$ | 0 | 0 |
| Turbulent kinetic energy | $\kappa$ | $\mu$ | $G - \rho\varepsilon$ | |
| Turbulent dissipation rate | $\varepsilon$ | $\mu$ | $C_{\varepsilon1} G \dfrac{\varepsilon}{k} - \rho C_{\varepsilon2} \dfrac{\varepsilon^2}{k}$ | |

**Table 2 Sutherland-law viscosity parameters for gases [9]**

| Gas | $T_0$ (K) | $\mu_0$ ($N\cdot s/m^2$) | $S$ (K) | Temperature range for 2% error (K) |
|---|---|---|---|---|
| Air | | 1.716E-5 | 111 | 170-1900 |
| Ar | | 2.125E-5 | 144 | 120-1500 |
| $N_2$ | 273 | 1.664E-5 | 107 | 100-1500 |
| $H_2$ | | 8.411E-6 | 97 | 220-1100 |
| He | | 1.864E-4 | 79.4 | 200-1500 |

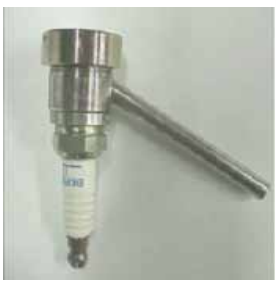 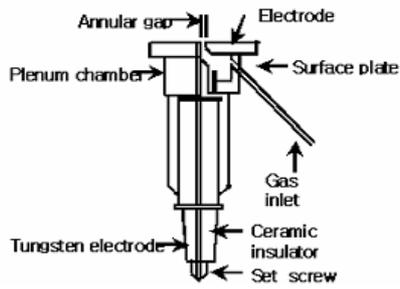 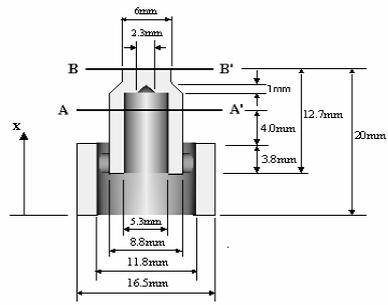

**Fig. 1 Schematic of a plasma torch.**

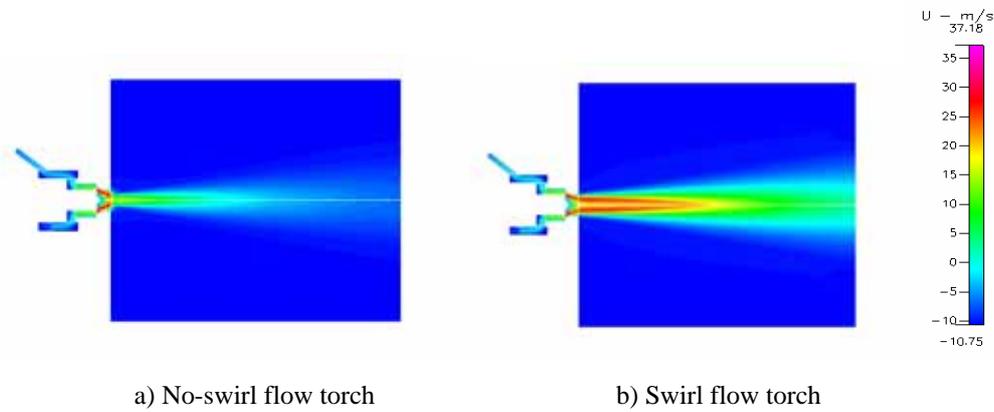

a) No-swirl flow torch  b) Swirl flow torch

**Fig. 2 Swirl effect on the velocity distribution in air (unit: m/s).**

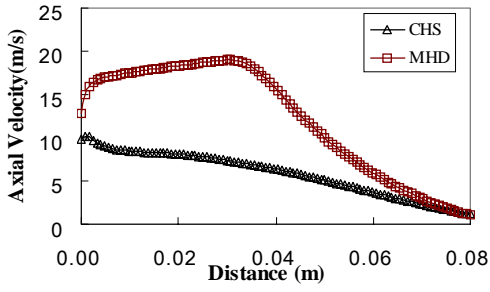
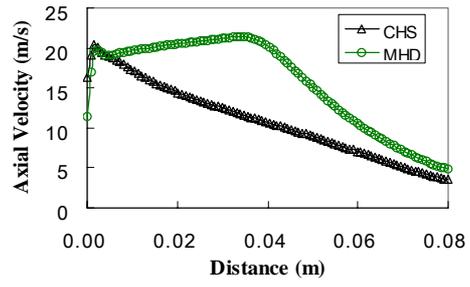

(a) Air

(b) Argon

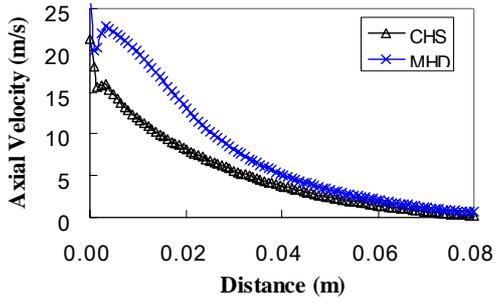
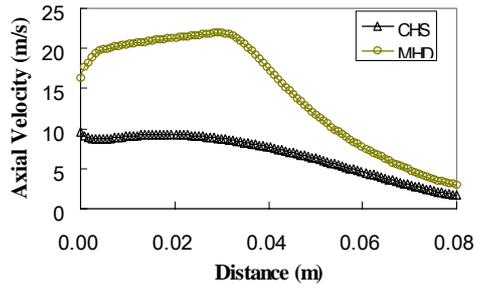

(c) Helium

(d) Nitrogen

**Fig. 3 Velocity distribution for different gases along the centerline (y=0).**

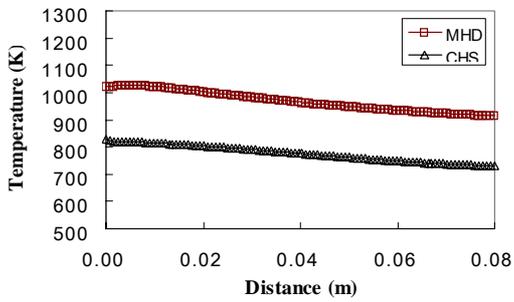
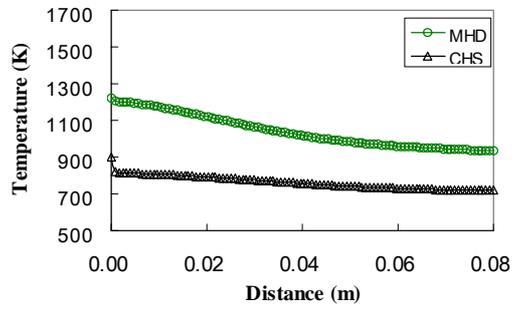

(a) Air

(b) Argon

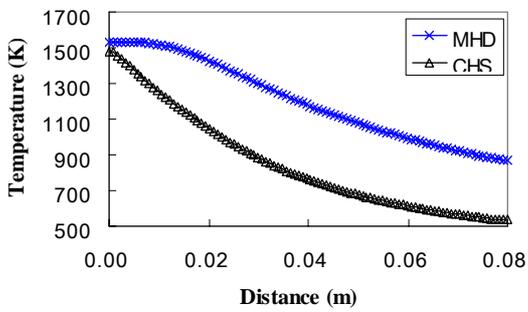
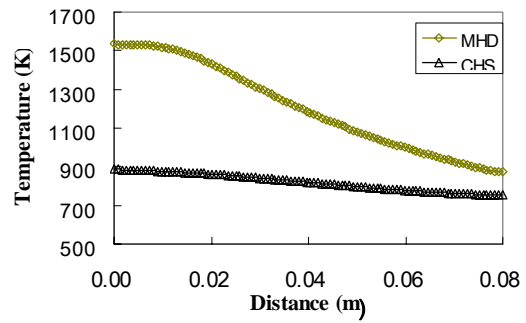

(c) Helium

(d) Nitrogen

**Fig. 4 Temperature distribution for different gases along the centerline (y=0).**